\documentclass[twocolumn,aps,prb,groupedaddress]{revtex4-1}
\usepackage{graphicx}
\usepackage{color}
\usepackage{amsmath}
\usepackage{enumitem}
\usepackage{amssymb}
\usepackage{hyperref}

\newcommand{\be}{\begin{equation}}
\newcommand{\ee}{\end{equation}}

\newcommand{\bea}{\begin{eqnarray}}
\newcommand{\eea}{\end{eqnarray}}

\renewcommand{\vec}[1]{{\bf #1}}

\begin{document}


\title{Tunable Quantum Hall Edge Conduction in Bilayer Graphene through Spin-Orbit Interaction}

\author{Jun Yong Khoo}
\affiliation{Department of Physics, Massachusetts Institute of Technology, 77 Massachusetts Avenue, Cambridge, Massachusetts 02139, USA}

\author{Leonid Levitov}
\affiliation{Department of Physics, Massachusetts Institute of Technology, 77 Massachusetts Avenue, Cambridge, Massachusetts 02139, USA}


\date{\today}

\begin{abstract}
Bilayer graphene, in the presence of a one-sided spin-orbit interaction (SOI) induced by a suitably chosen substrate, is predicted to exhibit unconventional Quantum Hall states. The new states arise due to strong SOI-induced splittings of the eight zeroth Landau levels, which are strongly layer-polarized, residing fully or partially on one of the two graphene layers. 
In particular, an Ising SOI in the meV scale is sufficient to invert the Landau level order between the $n=0$ and $n=1$ orbital levels under moderately weak magnetic fields $B \lesssim 10$\,T. Furthermore, when the Ising field opposes the $B$ field, the order of the spin-polarized levels can also be inverted. 
We show that, under these conditions, three different compensated electron-hole phases, with equal concentrations of electrons and holes, 
can occur at $\nu = 0$ filling. 
The three phases have distinct edge conductivity values. 
One of the phases is especially interesting, since its edge conduction can be turned on and off by switching the sign of the interlayer bias. 
\end{abstract}

\pacs{}

\maketitle


The effect of spin-orbit interaction (SOI) on the Landau levels (LLs) in graphene has been largely unexplored experimentally.
The main obstacle has been the extreme weakness of the intrinsic SOI, corresponding to spin splittings as small as 1 to 100\,$\mu$eV in graphene monolayer \cite{GSOI1,GSOIExpt,GSOIab3} and bilayer \cite{BLGSOI1,BLGSOI2,BLGSOI3}, which translates into splittings smaller than  the Zeeman energy $E_Z \sim 0.1 B$(T)\,meV even for relatively weak magnetic fields.
However, the situation has changed with the advent of graphene-based heterostructures. An SOI of 1 to 10\,meV has been interfacially-induced in graphene by transition-metal dichalcogenide substrates with strong SOI such as MoS$_2$, MoSe$_2$, WS$_2$ and WSe$_2$ \cite{BLGTMD,MLGWS2,MLGWS2b,MLGWS2WSe2,MLGMoS2WSe2,MLGMoSe2,MLGWS2MoS2} while an SOI of $\lesssim 100$\,meV was achieved at the graphene-Ni interface through Au intercalation \cite{GiantSOI}. These developments have opened the door to probing SOI-based physics in graphene.

It is particularly interesting to study how the LLs of (Bernal-stacked) bilayer graphene are modified by interfacially-induced SOI. Apart from having spin and valley degrees of freedom, the low-energy carriers in bilayer graphene are sensitive to the potential difference between the two graphene layers. These properties collectively give rise to a gate-tunable single-particle LL spectrum \cite{LLBLGbiased,LLmultiLGbiased}. 
After including the effects of electron-electron interactions, a rich phase diagram which hosts gate-tunable phase transitions is obtained\cite{QHFM1,QHFM2,QHFM3} that can be directly probed experimentally \cite{FracFill2,AFYoung,IntFracFill}. 
This gate-tunability is therefore expected to provide a means to probe the effects of interfacially-induced SOI on the LLs.

Here we consider the zeroth LLs of bilayer graphene, the set of eight-fold nearly degenerate lowest energy bands. The states of the zeroth LLs belonging to different valleys are strongly localized on different layers. This has a number of interesting implications. First, the valley degeneracy can be lifted by introducing layer-asymmetry to the system \cite{LLmultiLGbiased}. This can be achieved by applying an interlayer bias or by constructing an inversion-asymmetric heterostructure. Second, the layer asymmetric effects will be most noticeable in the zeroth LLs. 
Motivated by these observations, and in departure from previous treatments which considered SOI of equal strength for both layers \cite{BLGSOItheory,TLGSOI}, we shall focus on bilayer graphene with a \textit{layer-specific} SOI.
Recent experimental progress has led to a better understanding of the bilayer graphene zeroth LLs at various integer \cite{IntFill2,IntFill3,IntFill5,IntFracFill,AFYoung} and fractional fillings \cite{FracFill1,FracFill2,IntFracFill}, as well as to the recently demonstrated artificial SOI enhancement in graphene \cite{BLGTMD,MLGWS2,MLGWS2b,MLGWS2WSe2,MLGMoS2WSe2,MLGMoSe2,MLGWS2MoS2,GiantSOI}. We thus find ourselves in the opportune moment to investigate the novel valley-asymmetric effects on the zeroth LLs due to substrate-induced SOI.

To this end, here we analyze the single-particle LL spectrum of bilayer graphene with a layer-specific SOI of both the Ising and Rashba types. We highlight several interesting features that arise already at the non-interacting level. Some of these features are expected to remain robust in the presence of interactions. In particular, an Ising SOI $\lambda $ at the meV scale is strong enough to significantly change the zeroth LL spectrum. In contrast, Rashba SOI is of an off-diagonal character, and thus its effect is small even at values as large as $\lambda _{\rm R} \sim 15$\,meV. As a result, the energy ordering of the zeroth LL states is essentially determined by the competition between the layer-asymmetric Ising splitting and the Zeeman as well as orbital splittings.
In particular, the orbital and spin order inversions occur at relatively weak and moderate $B$ field values, respectively, as we discuss in detail below.

We predict three compensated electron-hole phases in this system, one of which is a conventional phase, whereas the other two phases arise due to the layer-asymmetric nature of the Ising SOI.
These three phases occur at the $\nu = 0 $ filling and when the interlayer bias $u$ is moderately large so that one layer becomes electron-doped while the other becomes hole-doped by the same amount.
In the absence of SOI, there is only one compensated electron-hole phase at $\nu=0$, which is expected to host helical edge modes with opposite chiralities and spin polarizations in each layer~\cite{AFYoung}.
Spin wavefunctions of these edge modes are orthogonal, forbidding
interlayer tunneling processes, and thereby protecting the edge states from backscattering. The expected edge conductance in this phase is therefore quantized at $2e^2/h$~\cite{helical}. 

\begin{widetext}

\begin{figure}[t]
\includegraphics[width=1.0\textwidth]{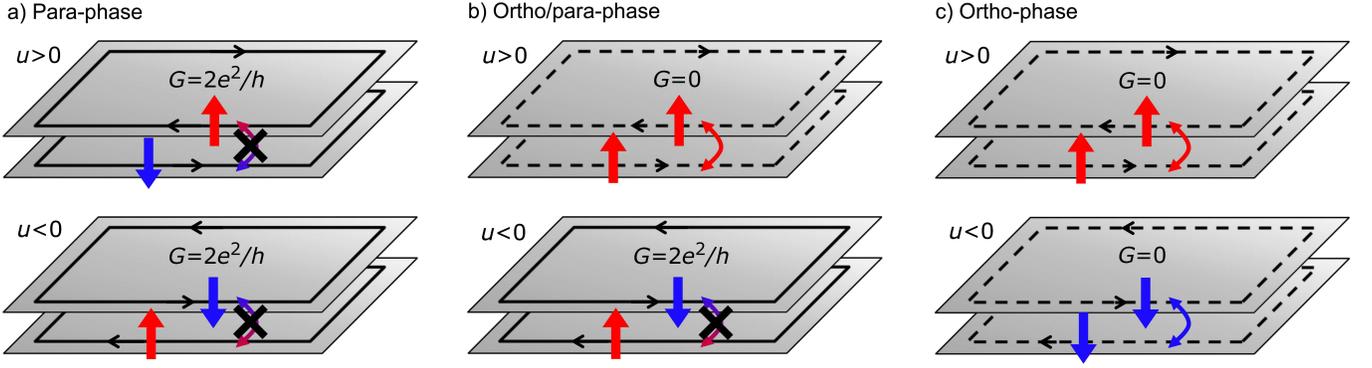}
\caption{\label{cartoonpic} Edge state configurations for a) para-phase, b) ortho/para-phase and c) ortho-phase. Edge conduction is reduced due to backscattering between the counterpropagating modes with equal spin polarization. Backscattering between the modes with opposite spins cannot occur because of the orthogonality of the spin wavefunctions, leading to quantized edge conduction. In the phase b), spin polarization of one of the edge modes can be reversed by a transverse electric field, giving rise to a switchable edge conduction.}
\end{figure}

\end{widetext}

Introducing a layer-specific Ising SOI can invert the energy ordering of the spin-polarized zeroth LLs set by the Zeeman splitting. This inversion occurs only within the zeroth LLs of the corresponding valley, and therefore requires that the Ising splitting at that valley be opposite in sign to the Zeeman splitting, and dominate over it. In this case, the corresponding compensated electron-hole phase will have helical edge modes with the same spin polarization. Backscattering processes are now allowed, so that the edge conductivity of this new compensated electron-hole phase is expected to be suppressed compared to the $2e^2/h$ value.

Increasing $B$ eventually restores the energy ordering so that we recover the compensated electron-hole phase equivalent to that in the system without SOI. However, the magnitude of Ising splitting is somewhat larger for the more strongly layer-polarized $n=0$ LLs as compared to the $n=1$ LLs. As a result, in the presence of moderately large $B$ fields only the ordering between the $n=1$ spin-polarized levels is restored while that of the $n=0$ remains inverted. 
Therefore, the transition between the two phases, which are dominated by the spin-orbital and Zeeman interactions, occurs via a third intermediate phase. This intermediate phase has edge modes with the same spin polarization for positive interlayer bias, but with opposite spin polarization for negative interlayer bias.

There are therefore a total of three different compensated electron-hole phases distinguished by whether their edge modes have the same or opposite spin polarizations for the cases of positive and negative interlayer bias, as illustrated in Fig.\ref{cartoonpic}. We denote the three phases as the `ortho-phase', `para-phase' and `ortho/para-phase', in analogy with the orthohydrogen and parahydrogen molecules. The intermediate ortho/para-phase is particularly interesting because it hosts an edge conductivity that can be turned on or off by switching the sign of the interlayer bias.

The low-energy states of bilayer graphene near the Dirac points (the $K_+$ and $K_-$ valleys) can be modeled by an effective Hamiltonian, expressed in the $(A1, B1, A2, B2)$ basis as \cite{epptyBLG,AFYoung,TBpara} 
\begin{eqnarray}\label{eq.HBLG0}
&&H_0^{B=0} = \left( \begin{array}{cccc}
\frac{u}{2} & v_0 \pi ^{\dagger} & -v_4 \pi ^{\dagger} & 0 \\
v_0 \pi & \frac{u}{2} + \Delta ' & \gamma _1 & -v_4 \pi ^{\dagger} \\
-v_4 \pi & \gamma _1 & -\frac{u}{2} + \Delta ' & v_0 \pi ^{\dagger} \\
0 & -v_4 \pi & v_0 \pi & -\frac{u}{2}
\end{array} \right), \\
&&\pi = \hbar (\xi k_x + i k_y), \quad \pi ^{\dagger} = \hbar (\xi k_x - i k_y),\quad v_{0,4} = \frac{\sqrt{3}a}{2\hbar} \gamma _{0,4}. \nonumber
\end{eqnarray}
Here, $a = 2.46\,\text{\AA}$ is the monolayer graphene lattice constant, the sign factor $\xi = \pm 1$ serves as the valley index corresponding to the valley wavevectors $\vec{K}_{\pm} = (\pm \frac{4\pi}{3a},0)$. The wavevector $\vec{k} = (k_x,k_y)$ is measured relative to $\vec{K}_{\pm}$. The hopping parameters are denoted by: $\gamma _0 = 2.61$\,eV for the intralayer nearest neighbor hopping, $\gamma _1 = 0.361$\,eV for the interlayer coupling between orbitals on the dimer sites B1 and A2, and $\gamma _4 = 0.138$\,eV for the interlayer coupling between dimer and non-dimer orbitals A1 and A2 or B1 and B2. The parameter $\Delta ' = 0.015$\,eV describes the energy difference between dimer and non-dimer sites. 
The interlayer bias is given by $u = V_2 - V_1$ where $V_i$ is the potential on layer $i=1,2$.

We model the interfacially-induced SOI by introducing a layer-specific spin-orbit Hamiltonian to the monolayer subspace of the system \cite{L1SOI}, described by the components $A1$ and $B1$ of Eq.\eqref{eq.HBLG0}:
\begin{eqnarray}\label{eq.dHSO}
&&\delta H_{SO}^{(1)} = \delta H_{\rm Ising} + \delta H_{\rm R}, \\
&&\delta H_{\rm Ising} = \frac{\lambda}{2}\xi s_z, \quad  \delta H_{\rm R} = \frac{\lambda _{\text R}}{2} (\xi \sigma _x s_y - \sigma _y s_x ), \nonumber
\end{eqnarray}
where $s_i$ and $\sigma _i$ are the Pauli matrices corresponding to the spin and A/B sublattice degrees of freedom respectively.  Under time reversal, the spin variables $s_i$, as well as the valley index $\xi$, change sign, whereas the sublattices $A$ and $B$ are not interchanged. The SOI Hamiltonian, Eq.\eqref{eq.dHSO},  is therefore invariant under time reversal. However, it is not invariant under inversion.

Indeed, our interfacial SOI interaction is distinct in its symmetry properties from the intrinsic SOI for graphene monolayer analyzed by Kane and Mele \cite{QSHG}. Both of the SOI terms in Eq.\eqref{eq.dHSO} are extrinsic, i.e. they are allowed by symmetry only because of the presence of the transition metal dichalcogenide substrate \cite{GonTMDtheory}. In particular, the Ising term $\delta H_{\rm Ising}$ is of the same form as the intrinsic Ising SOI of transition metal dichalcogenides with broken inversion symmetry \cite{TMDSOI1,TMDSOI2}. Thus, the interfacial  SOI induced in the graphene monolayer (and hence in the bilayer) also breaks the inversion symmetry. 
Likewise, the term $\delta H_{\rm R}$, which has the standard low-energy form of Rashba SOI in graphene subject to a transverse electric field at the substrate/graphene interface~\cite{QSHG}, also breaks the inversion symmetry.

We neglect the small intrinsic SOI terms of bilayer graphene, as well as Rashba SOI terms generated by the transverse electric field due to the interlayer potential between the graphene layers. Spin splittings due to these effects have been estimated to be in the range of 1 to 100~\rm{$\mu$}eV \cite{BLGSOI1,BLGSOI2,BLGSOI3}, which are much smaller than those arising from the interfacially-induced SOI in the meV range \cite{BLGTMD,MLGWS2,MLGWS2b,MLGWS2WSe2,MLGMoS2WSe2,MLGMoSe2,MLGWS2MoS2}.

We introduce a perpendicular magnetic field $\vec{B} = B \hat{z}$ via the usual replacement of $k_i$ with $q_i=k_i-\frac{e}{\hbar}A_i$ ($i=x,y$) where $\vec{A}=(A_x,A_y)$ is the vector potential, $\vec{B} = \vec{\nabla} \times \vec{A}$. We construct the magnetic ladder operators,
\be\label{eq.aadag}
\hat{a} = \frac{l_B}{\sqrt{2}} (q_x + i q_y), \quad \hat{a}^{\dagger} = \frac{l_B}{\sqrt{2}} (q_x - i q_y), \quad
\ee
which satisfy $\left[ \hat{a} , \hat{a}^{\dagger} \right] = 1$, 
where $l_B = \sqrt{\frac{\hbar}{eB}}$ is the magnetic length. Substituting these quantities into the full Hamiltonian $H = H_0^B + \delta H_{SO}^{(1)} $ of the system, we perform matrix diagonalization to solve for the LLs (See Appendix for detailed derivation of the zero Landau levels of bilayer graphene with layer-specific spin-orbit interaction). 

\begin{figure}[h]
\includegraphics[width=0.5\textwidth]{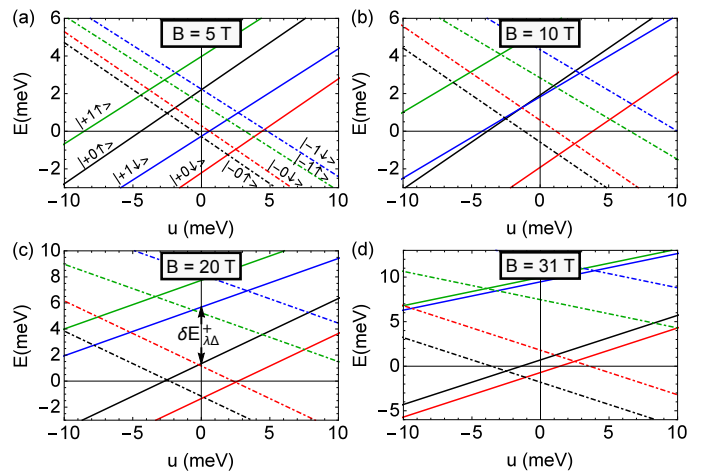}
\caption{Single-particle zeroth LL spectrum (including Zeeman splitting) as a function of interlayer bias $u$ for $\lambda = \lambda _{\rm R} = 5$\,meV at a) $B = 5$\,T, b) $B = 10$\,T, c) $B = 20$\,T, d) $B = 31$\,T. Different colors and linestyles are used to differentiate different LLs in each panel. The levels are labeled in a) by $|\xi n s_z \rangle$, the notation defined in Eq.\eqref{eq.LLlabel}. A reversal of the order of the levels $|+0\uparrow \rangle$ (black solid line) and $|+1\downarrow \rangle$ (blue solid line) occurs with increasing magnetic field, with the transition taking place at $B \approx 10$\,T shown in b).}
\label{lambdaLL}
\end{figure}

\begin{figure}[t]
\includegraphics[width=0.5\textwidth]{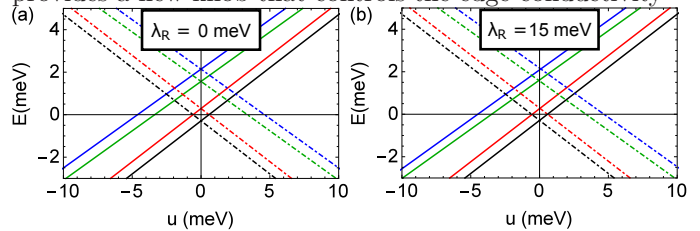}
\caption{\label{lambdaRLL}Illustration of the extreme smallness of the level shifts induced by the Rashba SOI $\lambda _{\rm R}$ in the zeroth LL, as compared to those induced by the Ising SOI $\lambda$ illustrated in Fig.~\ref{lambdaLL}. The spectrum found in the absence of SOI interaction remains essentially unchanged after adding a relatively large Rashba SOI: 
a) $\lambda =\lambda _{\rm R} = 0$\,meV and b) $\lambda = 0$, $\lambda _{\rm R} = 15$\,meV. Magnetic field is $B = 5$\,T in both cases. Labeling of Landau levels (color and linestyle) is the same as that in Fig.~\ref{lambdaLL}.
}
\end{figure}

In particular, the eigenstates of $H$ corresponding to the zeroth LLs are well parameterized by $|\xi n s_z \rangle$,
\be\label{eq.LLlabel}
H |\xi n s_z \rangle = E_{\xi , n, s_z} |\xi n s_z \rangle
\ee
Here $\xi=\pm1$ is the valley index, $n=0,1$ is the orbital LL index, and $s_z=\pm1$ is the out-of-plane spin polarization. 
To leading order in $u/\hbar \omega _c$, where $\omega _c$ is the cyclotron frequency, the corresponding single-particle energies are
\begin{eqnarray}\label{eq.EZLL}
&&E_{\xi , n, s_z} \simeq  -E_Z s_z + n \Delta _{10} + \frac{u}{2} \alpha _{\xi , n, s_z} + \xi \frac{\lambda}{2} \zeta _{1,\xi ,n}, \\
&&\Delta _{10} \approx  \hbar \omega _c\left(\frac{\Delta '}{\gamma _1} + 2 \frac{\gamma _4}{\gamma _0} \right), \quad \hbar \omega _c = \frac{3a^2\gamma _0^2}{2l_B^2 \gamma _1 }, \nonumber
\end{eqnarray}
where $E_Z = \mu _B B$ is the Zeeman splitting, $\alpha _{\xi , n, s_z}$ is the layer polarization, $\zeta _{1,\xi ,n}$ is the spin polarization on layer~1 and $\Delta _{10}$ is the orbital splitting. These leading order energy corrections already account for most of the features given by the exact solutions, which are shown in Fig.~\ref{lambdaLL} and Fig.~\ref{lambdaRLL} for typical values of $\lambda$, $\lambda _{\rm R}$ and selected values of $B$. The effects of the layer-specific SOI can be understood by contrasting Fig.~\ref{lambdaLL} and Fig.~\ref{lambdaRLL}(b) to Fig.~\ref{lambdaRLL}(a), which shows the zeroth LL spectrum in the absence of SOI ($\lambda = \lambda _{\rm R} = 0$). In what follows, we discuss some of these effects and their implications.

As we will see, the SOI-induced changes to the zeroth LL spectrum arise mainly due to the Ising SOI $\lambda$. This is illustrated by comparing the zeroth LL spectrum for $\lambda = \lambda _{\rm R} = 5$\,meV and $B = 5$\,T, shown in Fig.~\ref{lambdaLL}(a), to Fig.~\ref{lambdaRLL}(b), which shows the changes in the spectrum solely due to the $\lambda _{\rm R}$ coupling for the same $B$ field strength as in Fig.~\ref{lambdaLL}(a). 
In comparison to the Ising SOI, the effect of the Rashba SOI $\lambda _{\rm R}$ on the zeroth LLs is negligible and will not play a significant role in our analysis. A more detailed discussion of the reasons that effects due to $\lambda _{\rm R}$ are small can be found below (see the penultimate paragraph).

The Ising term $\lambda$ generates a valley-antisymmetric Zeeman-like splitting with opposite signs at the two valleys. 
Depending on the relative sign between $\lambda$ and $B$ at a given valley, this Ising field induces a splitting that either assists or counteracts the Zeeman splitting for a given valley. 
Importantly, because this Ising field is layer-specific, its splitting is directly proportional to the layer-1 polarisation of the state. For the zeroth LL states, the valley polarization is essentially in one-to-one correspondence with the layer polarization of the state. Consequently, the layer-specific Ising field influences the zeroth LLs in a valley-asymmetric fashion, whereby it only modifies the spectrum of the LL states in the $K_+$ valley (in our convention) and not those in the $K_-$ valley.

The most noticeable feature seen in Fig.~\ref{lambdaLL} is the evolution of the $|+0\uparrow \rangle$ (black solid line) and $|+1\downarrow \rangle$ (blue solid line) energy levels with increasing $B$ field from 5~T in Fig.~\ref{lambdaLL}(a) to 31~T in Fig.~\ref{lambdaLL}(d). From Eq.~\eqref{eq.EZLL}, we see that this is a direct consequence of the competition between the orbital splitting $\Delta _{10}$ and the Ising splitting $\lambda $. When the Ising splitting dominates in relatively weak $B$ fields, it changes the ordering between the more energetic $n=0$ and less energetic $n=1$ states in the $K_+$ valley. Consequently, at filling level $\nu = 2$, one of the $n=0$ and $n=1$ states are filled instead of both $n = 0$ states.

The onset of this inverted orbital ordering depends on the relative orientation between the Ising field and the external $B$ field. This inversion occurs when $\delta E_{\lambda \Delta} ^{\pm} < 0 $,
\begin{eqnarray}
\delta E_{\lambda \Delta} ^{+} &=& E_{+1\downarrow} - E_{+0\uparrow}, \quad \lambda> 0, \\
\delta E_{\lambda \Delta} ^{-} &=& E_{+1\uparrow} - E_{+0\downarrow}, \quad \lambda < 0.
\end{eqnarray}
which at $u = 0$ is approximately given by,
\begin{eqnarray}\label{eq.orbinv}
|\lambda| &\gtrsim & \Delta _{10} + \text{sgn}(\lambda) 2E_Z \\
&\simeq & \left\lbrace \begin{array}{c}
(0.381 + 0.116) B (T)\text{ meV}, \quad \lambda > 0 \\
(0.381 - 0.116) B (T)\text{ meV}, \quad \lambda < 0
\end{array} \nonumber, \right.
\end{eqnarray}
so that it occurs over a larger range of $B$ when the Ising and magnetic fields are aligned ($\lambda < 0$) than when they are anti-aligned ($\lambda > 0$). 

\begin{figure}[t]
\includegraphics[width=0.5\textwidth]{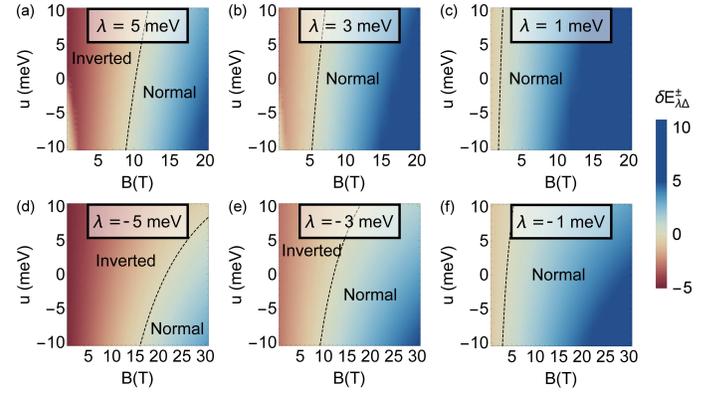}
\caption{\label{orbital} Phase diagrams in the $B$-$u$ plane showing how the regions of normal and inverted orbital ordering change with the Ising SOI $\lambda$ magnitude and sign. The panels a), b) and c) correspond to $\lambda > 0$ and normal/inverted ordering refers to the ordering between the levels $|+1\downarrow \rangle$ and $|+0\uparrow \rangle$: a) $\lambda = 5$\,meV, b) $\lambda = 3$\,meV, c) $\lambda = 1$\,meV. The panels d), e) and f) correspond to $\lambda < 0$ and normal/inverted ordering refers to the ordering between the levels $|+1\uparrow \rangle$ and $|+0\downarrow \rangle$ (see Eq.\eqref{eq.LLlabel} for notation): d) $\lambda = -5$\,meV, e) $\lambda = -3$\,meV, f) $\lambda = -1$\,meV. Dashed lines indicate the phase boundary along which $\delta E_{\lambda \Delta} ^{+} = 0$ [panels a)-c)], and $\delta E_{\lambda \Delta} ^{-}=0$ [panels d)-f)]. The energy difference $\delta E_{\lambda \Delta} ^{+}$ (in meV) is indicated in Fig.~\ref{lambdaLL}(c).}
\end{figure}

In Fig.~\ref{orbital}, we include the effects of the interlayer bias $u$ and map out the phase diagrams for several values of $\lambda= \pm 1, \pm 3, \pm 5$\,meV.  This serves to assist visualizing the region in the three-dimensional ($u, B, \lambda$) phase space in which orbital inversion occurs, i.e. when $\delta E_{\lambda \Delta} ^{\pm} < 0 $ is satisfied. 
For $\lambda > 0$, the ordering inversion occurs between $|+1\downarrow \rangle$ and $|+0\uparrow \rangle$ while for $\lambda < 0$, ordering inversion occurs between $|+1\uparrow \rangle$ and $|+0\downarrow \rangle$. The occurrence of inverted orbital ordering at the non-interacting level will likely lead to novel phases near the $\nu = 2$ filling when interaction effects are included.

In particular, two observations can be made from Fig.\ref{orbital}. First, the region in the $B-u$ phase space with orbital ordering inversion increases with the magnitude of $\lambda$. Second, these inversion regions are larger for $\lambda <0$ than for $\lambda >0$ of the same magnitude. These observations are consistent with what we have discussed above and are accounted for by Eq.~\eqref{eq.orbinv}.

Another interesting feature occurs at larger $B$ fields when orbital ordering of energy levels is restored. The occurrence of this feature also requires that the Ising splitting opposes the Zeeman splitting. Applying a moderately large interlayer potential $|u|$ dopes one layer into the electron band and the other into the hole band \cite{helical}. Three different compensated electron-hole phases, `ortho', `para' and `ortho/para', exist as a result of the competition between the Ising splitting and Zeeman splitting.
They are characterized by the alignment between the spin-polarization of their respective edge modes under positive and negative interlayer bias $u$ (see Fig.~\ref{cartoonpic}):
\begin{eqnarray}
\text{Para phase:}&& \quad \left\lbrace 
\begin{array}{c}
|\text{electron}, \uparrow \rangle \otimes |\text{hole}, \downarrow \rangle, \quad u>0 \\
|\text{electron}, \uparrow \rangle \otimes |\text{hole}, \downarrow \rangle, \quad u <0 
\end{array} \right.
, \nonumber \\
\text{Ortho-phase:}&& \quad \left\lbrace 
\begin{array}{c}
|\text{electron}, \uparrow \rangle \otimes |\text{hole}, \uparrow \rangle, \quad u >0 \\
|\text{electron}, \downarrow \rangle \otimes |\text{hole}, \downarrow \rangle, \quad u <0 
\end{array} \right.
, \nonumber \\
\text{Ortho/para-phase:}&& \quad \left\lbrace 
\begin{array}{c}
|\text{electron}, \uparrow \rangle \otimes |\text{hole}, \uparrow \rangle, \quad u >0 \\
|\text{electron}, \uparrow \rangle \otimes |\text{hole}, \downarrow \rangle, \quad u <0 
\end{array} \right.
. \nonumber \\
\end{eqnarray}

When the Ising field is absent or when the Zeeman splitting dominates, the compensated electron-hole phase is in the para-phase, which corresponds to `phase III' in Ref.~\onlinecite{AFYoung}. In this phase, the filled hole and electron bands residing on opposite layers have opposite spin-polarization. When the substrate-induced Ising splitting dominates ($\lambda \gg 2E_Z \simeq 0.116 B (T)$\,meV), the ordering of the spin-polarized states localized on the layer nearer to the substrate is now reversed. The corresponding compensated electron-hole phase is in the ortho-phase, in which case the filled electron band has the same spin-orientation as the filled hole band.

Unlike the para-phase, the ortho-phase is overall spin-neutral. In addition, having filled electron and hole bands with the same spin polarization means that their corresponding helical edge modes do not have protection from backscattering, unlike those of the para-phase.
The edge conductivity of the ortho-phase is therefore expected to be strongly suppressed compared to that of the para-phase.

The mixed ortho/para-phase occurs at moderate $B$ fields when the Ising and Zeeman splittings are comparable. 
Because the $|+0 s_z \rangle$ states are more strongly polarized on layer 1, they experience a stronger substrate-induced Ising field compared to the $|+1 s_z \rangle$ states. The Ising splitting between the $|+0 s_z \rangle$ states is therefore slightly larger than that between the $|+1 s_z \rangle$ states, so that they do not necessarily have the same spin ordering for a given value of $B$. This can be seen from the larger spin splitting between the $|+0 s_z \rangle$ states compared to that between the $|+1 s_z \rangle$ states in Fig.~\ref{lambdaLL}(d). Consequently, the para-phase to ortho-phase transition occurs at different values of $B$ when $u >0$ and when $u < 0$. To describe this ordering, we define the following parameters
\begin{eqnarray}\label{eq.dEpm}
\delta E_{\lambda B} ^{+} &=& E_{+0\downarrow} - E_{+0\uparrow}, \quad u> 0, \\
\delta E_{\lambda B} ^{-} &=& E_{+1\downarrow} - E_{+1\uparrow}, \quad u< 0. \nonumber
\end{eqnarray}
The conditions for the different phases are then given by
\begin{eqnarray}\label{eq.phasecond}
\text{Para phase:}&& \quad \delta E_{\lambda B} ^{+}>0,\quad \delta E_{\lambda B} ^{-} > 0, \nonumber \\
\text{Ortho-phase:}&& \quad \delta E_{\lambda B} ^{+}<0,\quad \delta E_{\lambda B} ^{-} < 0, \nonumber \\
\text{Ortho/para-phase:}&& \quad \delta E_{\lambda B} ^{+}<0,\quad  \delta E_{\lambda B} ^{-} > 0.
\end{eqnarray}

\begin{figure}[h]
\centering
\includegraphics[scale=0.5]{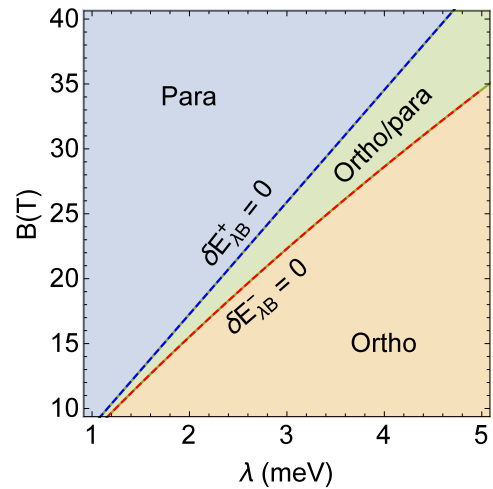}
\caption{\label{33barphasediag} Phase diagram in the $\lambda -B$ plane, showing the different regions in which the system is found in one of the three different compensated electron-hole phases: the para-phase, the ortho-phase, and the ortho/para-phase. The notation for the energy differences $\delta E_{\lambda B} ^{\pm}$ is given in Eq.\eqref{eq.dEpm}.}
\end{figure}

The orbital splitting gives rise to a mixed ortho/para-phase in the $\lambda - B$ plane (see Fig.~\ref{33barphasediag}), in which the system can be thought of as being in the ortho-phase for $u > 0$ and being in the para-phase for $u<0$. 
This mixed phase is particularly interesting because its edge conductivity can be switched on or off via switching the sign of the interlayer bias $u$.

While interaction effects are not included in this work, we expect the above discussion to remain qualitatively unchanged for $B \gtrsim 15$\,T. To understand why it is so, let us consider the $\lambda = 0$ case, in which the compensated electron-hole phase only exists in the para-phase (phase III in Ref.~\onlinecite{AFYoung}). 
The $\nu = 0$ phase diagram mapped out in Ref.~\onlinecite{AFYoung} suggests that the net effect of electron-electron interactions is to reduce the single-particle orbital splitting $\Delta_{10} \propto B$ which stabilizes the compensated electron-hole phase. In weak $B$ fields, orbital splitting is unable to overcome the interaction energy between the filled LLs on the same layer. 
In this case, the total energy of the system is minimized by one of two phases depending on the potential difference between the two layers. The ground state is in the layer-unpolarized canted antiferromagnetic phase (phase I in Ref.~\onlinecite{AFYoung}) when the potential on both layers are comparable. Otherwise, when the potential on one layer is much larger than the other, the ground state is in the completely layer-polarized, spin-neutral phase (phase II in Ref.~\onlinecite{AFYoung}).
The compensated electron-hole phase does not exist as a ground state of the system in weak $B$ fields. 
On the other hand, when $B \gtrsim 15$\,T, $\Delta_{10}$ is sufficiently large and can exceed the interaction energy between the filled LLs on the same layer. In this case, the physics is qualitatively captured by the single-particle picture. At moderately large values of $u$, the energy of the $|-1\uparrow \rangle$ state becomes lower than that of the $|+0\downarrow \rangle$ state (see Fig.\ref{lambdaRLL}(a)). The total energy of the system is therefore minimized by filling three LLs on one layer and one LL on the other -- i.e. the compensated electron-hole phase.

The novel phases, ortho and ortho/para, predicted in this work, arise due to the competition between the Zeeman splitting and the layer-specific Ising SOI splitting. This competition gives rise to the normal and inverted spin-ordering between LLs of the same orbital number $n$ and is therefore independent of the orbital splitting. We therefore do not expect the electron-electron interactions to have a big effect on the spin ordering, since they mainly affect the orbital splitting and, through that, may alter the effective interlayer potential. This effect, however, will have no direct impact on the competition between the SOI and Zeeman interactions that govern spin ordering. 
Therefore, so long as $B \gtrsim 15$\,T, the $\nu = 0$ ground state of the interacting system will be found in one of the three compensated electron-hole phases. This means that for an Ising field strength of $\lambda \gtrsim 2$\,meV, all three phases are expected to be accessible even when the interaction effects are included, since the transitions between the different phases occur at $B \gtrsim 15$ (see Fig.~\ref{33barphasediag}).

Following this reasoning, we expect that for these moderately large values of $B$, the $\nu = 0$ layer-unpolarized canted antiferromagnetic phase discussed in Ref. \onlinecite{AFYoung} is likely to remain unchanged in the presence of a strong SOI substrate.
However, this also means that the $\nu = 0$ phase diagram may change at weaker $B$ fields because of the inverted orbital ordering (see Fig.~\ref{orbital}). A more detailed study including interaction effects is required to map out the phase diagram in this regime.

An immediate consequence of these three different phases is the difference between their two-terminal conductances $G$. The para-phase conductance, as discussed earlier, is expected to take a quantized value $G_{\text{para}} = 2e^2/h$ for both positive and negative values of $u$ (each edge contributes $e^2/h$ in parallel). For the ortho-phase, backscattering between the counterpropagating edge states is allowed. 
This causes the two terminal conductance to decay exponentially from $2e^2/h$ to 0  as the sample dimension increases. Therefore, for sufficiently large samples, we expect $G_{\text{ortho}} \simeq 0$ for both positive and negative values of $u$. Finally, we expect the mixed ortho/para-phase to have a conductance that is gate-tunable -- $G_{\text{ortho/para}}(u > 0) \simeq 0$ and $G_{\text{ortho/para}}(u < 0) = 2e^2/h$.

Finally, we comment on the effects of Rashba SOI $\lambda _{\rm R}$, which are important at high LLs but are negligible at the zeroth LL. The smallness of the Rashba SOI $\lambda _{\rm R}$ for the zeroth LL is illustrated in Fig.\ref{lambdaRLL}. 
Indeed, by comparing the zeroth LL spectra shown in Fig.~\ref{lambdaRLL}(a) to those in Fig.~\ref{lambdaRLL}(b)($\lambda = 0$, $\lambda _{\rm R} = 15$\,meV), it is evident that the effects of the Rashba SOI are strongly suppressed even at values of $\lambda _{\rm R}$ as large as $15$\,meV. 
It gives rise to an energy correction $\approx 10^{-2}$\,meV, which is comparable to the Zeeman energy at $B = 1$\,T, but quickly becomes negligible at larger field strengths ($B \gtrsim 5$\,T).
It is therefore justified to ignore the correction due to $\lambda _{\rm R}$ at leading order, which was done in Eq.~\eqref{eq.EZLL}.
The physical reason for this smallness is as follows.
The matrices $\sigma_i$ in the Rashba term generate $A1$-$B1$ couplings, which mix the zeroth LL state $|+,0,\downarrow \rangle = |A1\downarrow , 0\rangle $ with the dimer states at relatively high energies $\pm \gamma_1$. 
As a result, the $\lambda _{\rm R}$-dependent corrections scale as $\delta \propto \lambda _{\rm R} \left(\frac{\lambda _{\rm R}}{\gamma _1}\right)$.
Furthermore, because the high-energy states $\pm \gamma_1$ are particle-hole symmetric, their contributions cancel out at the lowest order. 
The $\lambda _{\rm R}$-dependent corrections survive only at the next order, giving a small contribution to the level shifts of the order $\delta \propto \lambda _{\rm R} \left(\frac{\lambda _{\rm R}}{\gamma _1}\right)^2$.

In summary, the layer-specific SOI adds some unique features to the bilayer graphene single-particle zeroth LL spectrum. Those include the occurrence of an interesting pattern of orbital and spin inversions in the LL energy ordering. In contrast to the SOI-induced splittings of higher LLs, which are dominated by the Rashba SOI, the zeroth LL splittings are dominated by the Ising SOI. 
Furthermore, the states with spin inversion give rise to novel compensated electron-hole ortho- and ortho/para-phases with a unique set of edge modes and gate-tunable edge conduction.
While we anticipate the discovery of other novel phases when electron-electron interaction effects are taken into account, the compensated electron-hole ortho-phase and ortho/para-phase predicted here are expected to be robust to electron-electron interactions when $B \gtrsim 15$\,T. The ortho/para-phase in particular provides a new knob that controls the edge-conductivity of quantum Hall systems. 
The gate tunability enables a field-effect transistor-like behavior of the edge states, a property that can lead to interesting applications of graphene-on-transition metal dichalcogenide heterostructures.

J. K. acknowledges support by the National Science Scholarship from the Agency for Science, Technology and Research (A*STAR). This work was supported, in part, by the STC Center for Integrated Quantum Materials, NSF Grant No. DMR-1231319; and by Army Research Office Grant W911NF-18-1-0116.


%


\appendix

\section{Derivation of Zero Landau Levels}

In this appendix, we provide a derivation of the zeroth Landau level (LL) spectrum of BLG with layer-specific SOI. 
We follow the convention given in Ref.~\cite{epptyBLG} to describe the spin-degenerate, low-energy effective Hamiltonian of bernal-stacked bilayer graphene including the hopping parameters $\gamma _0, \gamma _1, \gamma _4, \Delta '$ as per Ref.~\cite{AFYoung}. The effective Hamiltonian is given in the main text Eq.~\eqref{eq.HBLG0}
. The monolayer SOI Hamiltonian Eq.~2 
 can be written explicitly as
\begin{eqnarray}
\delta H_{SO}^{(1)} 
&=& \left( \begin{array}{cccc}
\frac{\lambda}{2}\xi s_z & \frac{\lambda _R}{2} (\xi s_y + i s_x) & 0 & 0 \\
\frac{\lambda _R}{2} (\xi s_y - i s_x) & \frac{\lambda}{2}\xi s_z & 0 & 0 \\
0 & 0 & 0 & 0 \\
0 & 0 & 0 & 0 
\end{array}
\right), \nonumber \\
\end{eqnarray}
and will be added to the layer-1 subspace of the Hamiltonian. Following the main text, we then introduce a perpendicular magnetic field and construct Landau level creation and annihilation operators $\hat{a}$ and $\hat{a}^{\dagger}$. Defining these operators as in Eq.~\eqref{eq.aadag}
 gives rise to the following valley-specific transformations,

\begin{eqnarray}
\xi &=& +1: \left\lbrace \begin{array}{c}
\pi \rightarrow \frac{\sqrt{2}\hbar}{l_B} \hat{a} \\
\pi ^{\dagger} \rightarrow \frac{\sqrt{2}\hbar}{l_B} \hat{a}^{\dagger}
\end{array} \right. \\
\xi &=& -1: \left\lbrace \begin{array}{c}
\pi \rightarrow -\frac{\sqrt{2}\hbar}{l_B} \hat{a}^{\dagger} \\
\pi ^{\dagger} \rightarrow -\frac{\sqrt{2}\hbar}{l_B} \hat{a} 
\end{array} \right.
\end{eqnarray}
which we can substitute directly into the full Hamiltonian of the system $H = H_0 + \delta H_{SO}^{(1)}$, and write the following valley-specific Hamiltonians in the $(A1\uparrow,A1\downarrow,B1\uparrow,B1\downarrow,A2\uparrow,A2\downarrow,B2\uparrow,B2\downarrow)$ basis,
\begin{widetext}
\begin{eqnarray}
H_+ &=& \hbar \omega_0 \left( \begin{array}{cccccccc}
\frac{1}{2}(\bar{u} + \bar{\lambda}) & 0 & \hat{a}^{\dagger} & 0 & -\frac{\gamma_4}{\gamma_0} \hat{a}^{\dagger} & 0 & 0 & 0 \\
0 & \frac{1}{2}(\bar{u} - \bar{\lambda}) & i \bar{\lambda}_R & \hat{a}^{\dagger} & 0 & -\frac{\gamma_4}{\gamma_0} \hat{a}^{\dagger} & 0 & 0 \\
\hat{a} & -i \bar{\lambda}_R & \frac{1}{2}(\bar{u} + \bar{\lambda}) +\bar{\Delta} ' & 0 & \bar{\gamma}_1 & 0 & -\frac{\gamma_4}{\gamma_0} \hat{a}^{\dagger} & 0 \\
0 & \hat{a} & 0 & \frac{1}{2}(\bar{u} - \bar{\lambda}) +\bar{\Delta} ' & 0 & \bar{\gamma}_1 & 0 & -\frac{\gamma_4}{\gamma_0} \hat{a}^{\dagger} \\
-\frac{\gamma_4}{\gamma_0} \hat{a} & 0 & \bar{\gamma}_1 & 0 & -\frac{1}{2}\bar{u} +\bar{\Delta} ' & 0 & \hat{a}^{\dagger} & 0\\
0 & -\frac{\gamma_4}{\gamma_0} \hat{a} & 0 & \bar{\gamma}_1 & 0 & -\frac{1}{2}\bar{u} +\bar{\Delta} ' & 0 & \hat{a}^{\dagger} \\
0 & 0 & -\frac{\gamma_4}{\gamma_0} \hat{a} & 0 & \hat{a} & 0 & -\frac{1}{2}\bar{u} & 0 \\
0 & 0 & 0 & -\frac{\gamma_4}{\gamma_0} \hat{a} & 0 & \hat{a} & 0 & -\frac{1}{2}\bar{u}
\end{array}
\right),\label{eq.blgsop} \\
H_- &=& \hbar \omega_0 \left( \begin{array}{cccccccc}
\frac{1}{2}(\bar{u} - \bar{\lambda}) & 0 & -\hat{a} & i \bar{\lambda}_R & \frac{\gamma_4}{\gamma_0} \hat{a} & 0 & 0 & 0 \\
0 & \frac{1}{2}(\bar{u} + \bar{\lambda}) & 0 & -\hat{a} & 0 & \frac{\gamma_4}{\gamma_0} \hat{a} & 0 & 0 \\
-\hat{a}^{\dagger} & 0 & \frac{1}{2}(\bar{u} - \bar{\lambda}) +\bar{\Delta} ' & 0 & \bar{\gamma}_1 & 0 & \frac{\gamma_4}{\gamma_0} \hat{a} & 0 \\
-i \bar{\lambda}_R & -\hat{a}^{\dagger} & 0 & \frac{1}{2}(\bar{u} + \bar{\lambda}) +\bar{\Delta} ' & 0 & \bar{\gamma}_1 & 0 & \frac{\gamma_4}{\gamma_0} \hat{a} \\
\frac{\gamma_4}{\gamma_0} \hat{a}^{\dagger} & 0 & \bar{\gamma}_1 & 0 & -\frac{1}{2}\bar{u} +\bar{\Delta} ' & 0 & -\hat{a} & 0\\
0 & \frac{\gamma_4}{\gamma_0} \hat{a}^{\dagger} & 0 & \bar{\gamma}_1 & 0 & -\frac{1}{2}\bar{u} +\bar{\Delta} ' & 0 & -\hat{a}\\
0 & 0 & \frac{\gamma_4}{\gamma_0} \hat{a}^{\dagger} & 0 & -\hat{a}^{\dagger} & 0 & -\frac{1}{2}\bar{u} & 0 \\
0 & 0 & 0 & \frac{\gamma_4}{\gamma_0} \hat{a}^{\dagger} & 0 & -\hat{a}^{\dagger} & 0 & -\frac{1}{2}\bar{u}
\end{array}
\right). \label{eq.blgsom}
\end{eqnarray}

\end{widetext}
For brevity, we introduce $\hbar \omega _0 = \hbar \frac{\sqrt{2}v_0}{l_B} = \sqrt{\frac{3}{2}}\frac{a\gamma _0}{l_B}$ and various barred quantities which are related to their unbarred counterparts via $x = \hbar \omega_0 \bar{x}$. 

The creation and annihilation operators act on the site-specific Landau level wavefunctions $|n \rangle$ in the usual sense, $\hat{a} |n \rangle = \sqrt{n}|n -1 \rangle$ and $\hat{a}^{\dagger} |n \rangle = \sqrt{n+1}|n+1 \rangle$ so that by considering the following valley-specific ansatz for $n\geq 3$,
\begin{widetext}
\begin{eqnarray}
|+, n, i \rangle &=& \left( c^+_{i,A1\uparrow}|n \rangle ,c^+_{i,A1\downarrow} |n -1\rangle , c^+_{i,B1\uparrow}|n-1 \rangle , c^+_{i,B1\downarrow}|n-2 \rangle ,
c^+_{i,A2\uparrow}|n-1 \rangle , c^+_{i,A2\downarrow}|n-2 \rangle , c^+_{i,B2\uparrow}|n-2 \rangle , c^+_{i,B2\downarrow}|n-3 \rangle
\right)^T, \nonumber \\ \\
|-, n, i \rangle &=& \left( c^-_{i,A1\uparrow}|n-2 \rangle ,c^-_{i,A1\downarrow} |n-3 \rangle , c^-_{i,B1\uparrow}|n-1 \rangle , c^-_{i,B1\downarrow}|n-2 \rangle ,
c^-_{i,A2\uparrow}|n-1 \rangle , c^-_{i,A2\downarrow}|n-2 \rangle , c^-_{i,B2\uparrow}|n \rangle , c^-_{i,B2\downarrow}|n-1 \rangle
\right)^T, \nonumber \\
\end{eqnarray}
\end{widetext}
the Landau levels are given by the eigenvalues and eigenstates of the valley-specific matrices,
\begin{widetext}
\begin{eqnarray}
H_{+,n\geq 3} &=& \hbar \omega_0 \left( \begin{array}{cccccccc}
\frac{1}{2}(\bar{u} + \bar{\lambda}) & 0 & \sqrt{n} & 0 & -\frac{\gamma_4}{\gamma_0} \sqrt{n} & 0 & 0 & 0 \\
0 & \frac{1}{2}(\bar{u} - \bar{\lambda}) & i \bar{\lambda}_R & \sqrt{n-1} & 0 & -\frac{\gamma_4}{\gamma_0} \sqrt{n-1} & 0 & 0 \\
\sqrt{n} & -i \bar{\lambda}_R & \frac{1}{2}(\bar{u} + \bar{\lambda}) +\bar{\Delta} ' & 0 & \bar{\gamma}_1 & 0 & -\frac{\gamma_4}{\gamma_0} \sqrt{n-1} & 0 \\
0 & \sqrt{n-1} & 0 & \frac{1}{2}(\bar{u} - \bar{\lambda}) +\bar{\Delta} ' & 0 & \bar{\gamma}_1 & 0 & -\frac{\gamma_4}{\gamma_0} \sqrt{n-2} \\
-\frac{\gamma_4}{\gamma_0} \sqrt{n} & 0 & \bar{\gamma}_1 & 0 & -\frac{1}{2}\bar{u} +\bar{\Delta} ' & 0 & \sqrt{n-1}& 0\\
0 & -\frac{\gamma_4}{\gamma_0} \sqrt{n-1} & 0 & \bar{\gamma}_1 & 0 & -\frac{1}{2}\bar{u} +\bar{\Delta} ' & 0 & \sqrt{n-2} \\
0 & 0 & -\frac{\gamma_4}{\gamma_0} \sqrt{n-1} & 0 & \sqrt{n-1} & 0 & -\frac{1}{2}\bar{u} & 0 \\
0 & 0 & 0 & -\frac{\gamma_4}{\gamma_0} \sqrt{n-2} & 0 & \sqrt{n-2} & 0 & -\frac{1}{2}\bar{u}
\end{array}
\right), \nonumber \\ 
\end{eqnarray}

\begin{eqnarray}
H_{-,n\geq 3} &=& \hbar \omega_0 \left( \begin{array}{cccccccc}
\frac{1}{2}(\bar{u} - \bar{\lambda}) & 0 & -\sqrt{n-1} & i \bar{\lambda}_R & \frac{\gamma_4}{\gamma_0} \sqrt{n-1} & 0 & 0 & 0 \\
0 & \frac{1}{2}(\bar{u} + \bar{\lambda}) & 0 & -\sqrt{n-2} & 0 & \frac{\gamma_4}{\gamma_0} \sqrt{n-2} & 0 & 0 \\
-\sqrt{n-1} & 0 & \frac{1}{2}(\bar{u} - \bar{\lambda}) +\bar{\Delta} ' & 0 & \bar{\gamma}_1 & 0 & \frac{\gamma_4}{\gamma_0} \sqrt{n} & 0 \\
-i \bar{\lambda}_R & -\sqrt{n-2} & 0 & \frac{1}{2}(\bar{u} + \bar{\lambda}) +\bar{\Delta} ' & 0 & \bar{\gamma}_1 & 0 & \frac{\gamma_4}{\gamma_0} \sqrt{n-1} \\
\frac{\gamma_4}{\gamma_0} \sqrt{n-1} & 0 & \bar{\gamma}_1 & 0 & -\frac{1}{2}\bar{u} +\bar{\Delta} ' & 0 & -\sqrt{n} & 0\\
0 & \frac{\gamma_4}{\gamma_0} \sqrt{n-2} & 0 & \bar{\gamma}_1 & 0 & -\frac{1}{2}\bar{u} +\bar{\Delta} ' & 0 & -\sqrt{n-1}\\
0 & 0 & \frac{\gamma_4}{\gamma_0} \sqrt{n} & 0 & -\sqrt{n} & 0 & -\frac{1}{2}\bar{u} & 0 \\
0 & 0 & 0 & \frac{\gamma_4}{\gamma_0} \sqrt{n-1} & 0 & -\sqrt{n-1} & 0 & -\frac{1}{2}\bar{u}
\end{array}
\right). \nonumber \\
\end{eqnarray}
\end{widetext}

By setting to zero all other energy parameters in Eq.~\eqref{eq.HBLG0} 
except $\gamma _0$ and $\gamma _1$, we recover the eight-fold degenerate (2 spin, 2 valley, 2 orbital) zeroth LLs states with zero energy, $|\xi , 0, s_z \rangle$ and $|\xi , 1, s_z \rangle$. By considering $H_{\pm,n=0,1,2}$ in this limit, we can understand how the zeroth LL degeneracy is lifted (i.e. how these states mix with the higher Landau levels and the terms responsible for the mixing). 
At $n = 0$, the solutions are immediately given,

\begin{eqnarray}
H_{+,0} &=& \frac{1}{2}(u + \lambda), \quad \text{eigenstate:} |+,0,\uparrow \rangle = |A1\uparrow,0 \rangle  \nonumber \\ \\ 
H_{-,0} &=& -\frac{1}{2}u, \quad \quad \text{eigenstate:} |-,0,\uparrow \rangle = |B2\uparrow,0 \rangle . \nonumber \\ 
\end{eqnarray}
The eigenstates and energies are exact. At $n=1$, we have
\begin{widetext}
\begin{eqnarray}
H_{+,1} &=& \hbar \omega_0 \left( \begin{array}{cccc}
\frac{1}{2}(\bar{u} + \bar{\lambda}) & 0 & 1 & -\frac{\gamma_4}{\gamma_0} \\
0 & \frac{1}{2}(\bar{u} - \bar{\lambda}) & i \bar{\lambda}_R & 0 \\
1 & -i \bar{\lambda}_R & \frac{1}{2}(\bar{u} + \bar{\lambda}) +\bar{\Delta} ' &  \bar{\gamma}_1 \\
-\frac{\gamma_4}{\gamma_0} & 0 & \bar{\gamma}_1 & -\frac{1}{2}\bar{u} +\bar{\Delta} ' \\
\end{array}
\right), \quad \text{basis:} \left( |A1\uparrow,1 \rangle , |A1\downarrow,0 \rangle, |B1\uparrow,0 \rangle, |A2\uparrow,0 \rangle \right) \nonumber \\ \\
%
H_{-,1} &=& \hbar \omega_0 \left( \begin{array}{cccc}
\frac{1}{2}(\bar{u} - \bar{\lambda}) +\bar{\Delta} ' & \bar{\gamma}_1 & \frac{\gamma_4}{\gamma_0} & 0 \\
\bar{\gamma}_1 & -\frac{1}{2}\bar{u} +\bar{\Delta} ' & -1 & 0\\
\frac{\gamma_4}{\gamma_0} & -1 & -\frac{1}{2}\bar{u} & 0 \\
0 & 0 & 0 & -\frac{1}{2}\bar{u}
\end{array}
\right), \quad \text{basis:} \left( |B1\uparrow,0 \rangle , |A2\uparrow,0 \rangle, |B2\uparrow,1 \rangle, |B2\downarrow,0 \rangle \right). \nonumber \\
\end{eqnarray}
\end{widetext}

Keeping only $\gamma _0$ and $\gamma _1$ non-trivial and setting the other parameters to 0, we find that the zeroth LL eigenstates are
\begin{eqnarray}
|+, 1, \uparrow \rangle _0 &=& \frac{1}{\sqrt{\bar{\gamma}_1^2 + 1}}\left(-\bar{\gamma}_1 |A1\uparrow, 1 \rangle + |A2\uparrow , 0 \rangle \right),  \nonumber \\ \\
|+, 0, \downarrow \rangle _0 &=& |A1\downarrow , 0\rangle ,  \nonumber \\ \\
|-, 1, \uparrow \rangle _0 &=& \frac{1}{\sqrt{\bar{\gamma}_1^2 + 1}}\left(|B1\uparrow , 0 \rangle + \bar{\gamma}_1 |B2\uparrow, 1 \rangle  \right),  \nonumber \\ \\
|-, 0, \downarrow \rangle _0 &=& |B2\downarrow , 0\rangle , \nonumber \\
\end{eqnarray}
With the above information, we can now discuss the effects due to the various parameters in the Hamiltonian as corrections which are justified in the strong field limit $\hbar \omega _0 \simeq 31$ meV$\sqrt{B (T)} \gg \frac{\gamma _4}{\gamma _0}, \Delta ', u, \lambda , \lambda _R$ but constrained to $\hbar \omega _0 \lesssim \gamma _1$ or equivalently $\bar{\gamma} _1 \gtrsim 1$. The effects from SOI are already discussed in the main text and here we include a short discussion on the effects of the non-SOI parameters for completeness. The interlayer bias $u$ shifts the energies by an amount that measures the layer polarization of the state as is expected, $+\frac{u}{2}$ if the state is completely polarized in layer 1 and $-\frac{u}{2}$ if it is completely polarized in layer 2. Similarly, the $\Delta '$ term gives rise to an energy shift proportional to the state's polarization on the $A2$ or $B1$ sites. Finally, $\gamma _4$ introduces mixing between the different sublattice components of the $|+, 1, \uparrow \rangle _0$ and $|-, 1, \uparrow \rangle _0$ respectively and shifts the state's energy by the difference between the amount of symmetric and anti-symmetric superposition of the sublattice components.

At $n=2$, we have
\begin{widetext}
\begin{eqnarray}
H_{+,2} &=& \hbar \omega_0 \left( \begin{array}{ccccccc}
\frac{1}{2}(\bar{u} + \bar{\lambda}) & 0 & \sqrt{2} & 0 & -\frac{\gamma_4}{\gamma_0} \sqrt{2} & 0 & 0 \\
0 & \frac{1}{2}(\bar{u} - \bar{\lambda}) & i \bar{\lambda}_R & 1 & 0 & -\frac{\gamma_4}{\gamma_0} & 0 \\
\sqrt{2} & -i \bar{\lambda}_R & \frac{1}{2}(\bar{u} + \bar{\lambda}) +\bar{\Delta} ' & 0 & \bar{\gamma}_1 & 0 & -\frac{\gamma_4}{\gamma_0} \\
0 & 1 & 0 & \frac{1}{2}(\bar{u} - \bar{\lambda}) +\bar{\Delta} ' & 0 & \bar{\gamma}_1 & 0 \\
-\frac{\gamma_4}{\gamma_0} \sqrt{2} & 0 & \bar{\gamma}_1 & 0 & -\frac{1}{2}\bar{u} +\bar{\Delta} ' & 0 & 1 \\
0 & -\frac{\gamma_4}{\gamma_0} & 0 & \bar{\gamma}_1 & 0 & -\frac{1}{2}\bar{u} +\bar{\Delta} ' & 0 \\
0 & 0 & -\frac{\gamma_4}{\gamma_0} & 0 & 1 & 0 & -\frac{1}{2}\bar{u}
\end{array}
\right), \nonumber \\ 
&\text{basis:} &\quad \left( |A1\uparrow,2 \rangle, |A1\downarrow,1 \rangle, |B1\uparrow,1 \rangle, |B1\downarrow,0 \rangle , |A2\uparrow,1 \rangle, |A2\downarrow,0 \rangle, |B2\uparrow,0 \rangle \right) \nonumber \\ \\
H_{-,2} &=& \hbar \omega_0 \left( \begin{array}{ccccccc}
\frac{1}{2}(\bar{u} - \bar{\lambda}) & -1 & i \bar{\lambda}_R & \frac{\gamma_4}{\gamma_0} & 0 & 0 & 0 \\
-1 & \frac{1}{2}(\bar{u} - \bar{\lambda}) +\bar{\Delta} ' & 0 & \bar{\gamma}_1 & 0 & \frac{\gamma_4}{\gamma_0} \sqrt{2} & 0 \\
-i \bar{\lambda}_R & 0 & \frac{1}{2}(\bar{u} + \bar{\lambda}) +\bar{\Delta} ' & 0 & \bar{\gamma}_1 & 0 & \frac{\gamma_4}{\gamma_0} \\
\frac{\gamma_4}{\gamma_0} & \bar{\gamma}_1 & 0 & -\frac{1}{2}\bar{u} +\bar{\Delta} ' & 0 & -\sqrt{2} & 0\\
0 & 0 & \bar{\gamma}_1 & 0 & -\frac{1}{2}\bar{u} +\bar{\Delta} ' & 0 & -1 \\
0 & \frac{\gamma_4}{\gamma_0} \sqrt{2} & 0 & -\sqrt{2} & 0 & -\frac{1}{2}\bar{u} & 0 \\
0 & 0 & \frac{\gamma_4}{\gamma_0} & 0 & -1 & 0 & -\frac{1}{2}\bar{u}
\end{array}
\right), \nonumber \\ 
&\text{basis:} &\quad \left( |A1\uparrow,0 \rangle, |B1\uparrow,1 \rangle, |B1\downarrow,0 \rangle , |A2\uparrow,1 \rangle, |A2\downarrow,0 \rangle, |B2\uparrow,2 \rangle , |B2\downarrow,1 \rangle \right). \nonumber \\ 
\end{eqnarray}

\end{widetext}
Once again, turning off all parameters except $\gamma _0$ and $\gamma _1$ allows us to recover the remaining two zeroth LL states:
\begin{eqnarray}
|+, 1, \downarrow \rangle _0 &=& \frac{1}{\sqrt{\bar{\gamma}_1^2 + 1}}\left(-\bar{\gamma}_1 |A1\downarrow, 1 \rangle + |A2\downarrow , 0 \rangle \right),  \nonumber \\ \\
|-, 1, \downarrow \rangle _0 &=& \frac{1}{\sqrt{\bar{\gamma}_1^2 + 1}}\left(|B1\downarrow , 0 \rangle + \bar{\gamma}_1 |B2\downarrow, 1 \rangle  \right),  \nonumber \\
\end{eqnarray}
The qualitative nature of how the various parameters affect these two zeroth LL states is similar to that of the $n=1$ case discussed earlier. 

Consistent with the above discussion, the single-particle energies of the zeroth LL to leading order in $u/\hbar \omega _c$ is given by main text Eq.~\eqref{eq.EZLL}
, and here we give the explicit expressions for the layer and spin polarizations that were omitted in the main text,
\begin{eqnarray}
\alpha _{\xi,n,s_z} &=& |c_{A1,s_z} |^2 + |c_{B1,s_z} |^2 - |c_{A2,s_z} |^2 - |c_{B2,s_z} |^2 \nonumber \\ \\ 
\zeta _{1,n, s_z} &=& s_z \left( |c_{A1,s_z} |^2 + |c_{B1,s_z} |^2 \right).
\end{eqnarray}

\end{document}